\documentclass[amssymb,aps,prli,preprint]{revtex4}
\usepackage{latexsym,amsmath,graphics,graphicx,exscale,colordvi,color}

\begin{document}
\title{Quantum well states and amplified spin-dependent Friedel oscillations in thin films}
%\title{Design of gigantic spin-dependent lateral electron focusing through quantum well states}
\author{Mohammed Bouhassoune}\email{m.bouhassoune@fz-juelich.de}
\author{Bernd Zimmermann}
\author{Phivos Mavropoulos}
\author{Daniel Wortmann}
\author{Peter H. Dederichs}
\author{Stefan Bl\"ugel}
\author{Samir Lounis}\email{s.lounis@fz-juelich.de}
\affiliation{Peter Gr\"unberg Institut \& Institute for Advanced Simulation, 
Forschungszentrum J\"ulich \& JARA, D-52425 J\"ulich,
Germany}

\begin{abstract}

Electrons mediate many of the interactions between atoms in a solid. Their propagation in a material determines its thermal, electrical, optical, 
magnetic and transport properties. Therefore, the constant energy contours characterizing the electrons, in particular the Fermi surface, have a prime impact on 
the behavior of materials. If anisotropic, the contours induce strong directional dependence at the nanoscale 
in the Friedel oscillations surrounding impurities. 
Here we report on giant anisotropic charge density oscillations focused along specific directions with strong spin-filtering  after 
scattering at an oxygen impurity embedded in the surface of a ferromagnetic thin film of Fe grown on W(001). 
Utilizing density functional theory, we demonstrate that by changing the thickness of the Fe films, we control quantum well states confined to two dimensions that manifest as multiple flat 
energy contours, impinging and tuning the strength of the induced charge oscillations which allow to detect the oxygen impurity at large distances ($\approx$ 50nm). 
\end{abstract} 
\maketitle
%\date{\today}

%\pacs{}

%\begin{multicols}{2}
%narrow text

Friedel oscillations arise from the charge screening of localized defects by the electron gas of metals.
The form and the magnitude of these oscillations do not only depend on the shape of the Fermi surface of 
the host but also on the nature of the defects. On the surface of noble metals, where the constant 
energy contour of the surface state is isotropic, 
lateral ring-like isotropic Friedel oscillations emanating from step edges and adatoms 
have been observed and predicted (see e.g.~\cite{Crommie1,Avouris1,Crommie2,Silly,Lounis2,Meier}). However, 
if the Fermi surface is anisotropic and 
bears flat regions where its curvature is small or vanishes, a focusing effect 
 with a strong 
directional bundling of group velocities can shape the electronic propagation 
at the nanoscale after scattering with defects~\cite{Weismann,Lounis,Avotina}. This has been shown for  Co impurities buried below 
copper surfaces by measurements 
performed with a scanning tunneling microscope (STM) combined with density functional theory (DFT)
 based simulations using the 
Korringa-Kohn-Rostoker Green function (KKR) method~\cite{Weismann,Lounis}. 
As a consequence, in real bulk materials Friedel oscillations are strongly anisotropic and can 
decay slower than $1/R^3$ contrary to the general wisdom stating that 
the confinement or dimensionality ($D$) of the electron gas  defines the decay of the charge oscillations ($~R^{-D}$). 
It was envisioned that a quantum mechanics induced phenomenon such as the focusing effect 
can be utilized in a nanosonar device that would allow to detect 
hidden buried impurities as well as buried nanostructures~\cite{Weismann}. 

The confinement of charge density oscillations leads to quantum well states (QWS), which underlie 
many intriguing phenomena, such as the modulated superconducting transition temperature in thin 
films~\cite{Guo}, the quantum oscillation of surface chemical reactivities~\cite{Ma}, the variation of Rashba 
splitting in topological insulators~\cite{Yang}, the oscillatory interlayer exchange coupling between 
magnetic films across a 
non-magnetic spacer and the giant magnetoresistance~\cite{Grunberg,Baibich,Parkin}. In particular, these 
QWS can get spin split in noble metals on magnetic 
substrates~\cite{Ortega,Carbone} as measured by photo-emission experiments. Similarly to thin films, the 
charge density oscillations govern magnetic interactions among single atoms and could act as a communication tool 
between 
nano-objects~\cite{Zhou,Khajetoorians1,Khajetoorians2,Ngo}. Thus, their impact in nanotechnologies is not negligible. 
For instance, these waves are important for the self-assembly of superstructures of adatoms 
on surfaces~\cite{Silly}, while the magnetic interactions (RKKY-type) lead to 
unforeseen magnetic ground states~\cite{Khajetoorians1,Pruser,Ngo,Stepanyuk1,Stepanyuk2}. Controlling and 
understanding the decay of these interactions, intimately related 
to the decay of the charge density oscillations, is decisive if these atoms are used as 
building blocks in nano-spintronic devices.

The ability to manipulate the energy contours, and in particular Fermi surfaces, with 
appropriate properties such as their shape and in particular their flatness, is desirable in 
order to control the magnitude of Friedel oscillations. In this work, we propose the use of exchange 
split QWS in magnetic thin films to design two-dimensional Fermi surfaces with multiple, layer-dependent, 
flat energy contours leading to an extraordinarily slow lateral decay of the charge density oscillation. 
Depending 
on the quantum number defining the QWS, the spatial localization of these states changes 
from layer to layer. This is different from the case of a surface state which is strongly 
confined on the surface layer. This idea concretized after the experimental observation of 
unexpected anisotropic charge density oscillations around the Fermi energy due to a non-magnetic 
impurity, most probably oxygen, implanted in the surface of a thin Fe film grown on W(001) surface~\cite{Bergmann}. These Friedel oscillations show fourfold in-plane symmetry and are focused 
along the diagonal $\langle 110 \rangle$ directions of the crystal surface as 
observed in the dI/dU map for 3 monolayers (MLs) Fe on W(001) at $U$ = -0.1 eV. Surprisingly, 
these oscillations extinguish if a thinner Fe film is considered, i.e. 2 instead of 3 MLs. To understand this experimental result, 
we utilize DFT to evaluate the 
charge density induced by the insertion of an oxygen impurity in the ferromagnetic surface of 2 
and 3 MLs Fe deposited on W(001) surface. After explaining the aforementioned experimental measurement, we assess the 
spin-nature of the measured Friedel oscillations and relate their focus along specific directions to the existence of 
QWS with flat constant energy contours that occur not only for 3 Fe MLs but also for thicker films. Moreover, we 
indicate the possible fundamental and technological implications of such lateral spin-density oscillations. 

\section*{Results}

\subsection*{Two and three monolayers of Fe on W(001)} 
Fig.~1 (a) represents a schematic of the STM measurements of the charge oscillations induced by an oxygen impurity embedded in the surface 
of Fe thin films on W(001) surface. The resulting highly anisotropic charge oscillations are illustrated in Fig.~1 (b)~\cite{Bergmann} while  
Figs.~1 (c, d, e, f) illustrate the spatial modulation of the induced charge density at the Fermi energy 
due to the oxygen atom calculated in the vacuum at a distance of 3.165 \AA~ 
above the Fe surface layer for different spin channels. Details about the theoretical methods based on DFT are given in the Methods section and the Supplementary Note 1. 
The Fe films are considered in their ferromagnetic ground states but we point out that 
contrary to 1 ML Fe/W(100) characterized by an antiferromagnetic ground state of the 
Fe ML~\cite{Spisak,Kubetzka}, thicker Fe films on the same substrate are 
ferromagnetic~\cite{vonBergmann_PRB}. 
The fourfold symmetry of the anisotropic 
oscillations observed in the experiment is obtained theoretically and is notably pronounced for 
the majority-spin channel of 3 MLs Fe and much stronger than those of its minority-spin channel or those obtained from 
both spin channels of 2 MLs Fe. The majority-spin density oscillations in the [110] 
direction have the highest amplitude and survive the longest away from the impurity. 
These results imply spin-filtering in that the spin-nature of the oscillations observed experimentally are mostly 
of majority-spin type. As explained below, this surprising behavior is induced by the confinement effects in the Fe films.

\subsection*{Quantum Well States} 
QWS of metallic films on substrates can be fully localized in the film. This naturally occurs for insulator or semiconductor substrates, since in the energy gap of the insulator there are no states available for interaction with the metallic-layer states. The states are then confined in the metallic layer, with small exponential tails in the vacuum region and in the substrate.  However, localized states can also occur for metallic overlayers on metallic substrates. When the three-dimensional Fermi surface of the substrate is projected on the two-dimensional surface Brillouin zone, there can be regions that are not covered by the projected Fermi surface. If the crystal momenta $\mathbf{k}$ of Bloch states with the projection $\mathbf{k_{\parallel}}=\mathbf{q}$ of the overlayer fall in such regions, these states cannot propagate into the substrate. Then QWS, confined in the overlayer, are formed at these particular regions of the surface Brillouin zone. Practically all metals have such $\mathbf{q}$-dependent gap region. They are rather large for the bcc metals W, Mo and Cr. The projected two-dimensional Fermi surface of W for the (001) Brillouin zone is shown in Fig.~2 (a). In addition to the four-fold symmetries one sees rather large gap-regions along the $\overline{\Gamma}\overline{\mathrm M}$ directions, while states exist around the origin $\mathbf{q}=0$ and along the x- and y-axis, i.e. along the $\overline{\Gamma}\overline{\mathrm X}$ directions.

When we compare the electronic structure of ferromagnetic Fe with the one of W (or Cr), we find that in the vicinity of $E_{\mathrm F}$ the minority-spin Fe bands are similar to the W bands, just shifted somewhat to higher energies. Therefore, also the Fermi surface of Fe minority electrons is similar to the one of W (or Cr) and the same is true for the (001)-projected two-dimensional Fermi surface (Fig.~2 (b)). There are also rather big $\mathbf{q}$-gaps along the $\overline{\Gamma}\overline{\mathrm M}$ directions and minority-spin quantum well states cannot occur in this $\mathbf{q}$-region. However, the situation is very different for the Fe majority-spin states. There the projected Fermi surface covers nearly the whole (001) surface Brillouin zone (Fig.~2 (c)). Consequently one expects that a thicker Fe-layer shows quantum well states confined for $q$-values within those gap regions of W.

Figure 3 shows the appearance of such quantum well states with increasing layer thickness. The two-dimensional Bloch spectral function of the Fe surface atoms at the Fermi energy is plotted in the upper figures for Fe layer thicknesses of 3, 5, 7, 9 and 11 MLs (see Supplementary Note 2 and Supplementary Figure 1 for  the case of an even number of MLs). As one sees, the gap regions are filled with an increasing number of confined states, one state for 3 MLs, 2 for 5 MLs and 5 for 11 MLs, etc. Most important is that these states show nearly no dispersion, i.e. they are straight lines perpendicular to the $\langle 110 \rangle$ diagonals. Therefore the group velocities of all these states are focused in the $\langle 110 \rangle$  directions. The lower pictures show the energy dependence of these states for $q$-values along  $\overline\Gamma \overline{\mathrm M}$, i.e. along the diagonal.

The majority of the states in Fig.~3 (lower panel) show a parabolic band structure. Moreover, by increasing the layer thickness by 2 MLs, one additional parabola is pushed down from higher energies below the Fermi level, in order to accommodate the additional number of electrons. For small $q$-values the parabolic states are damped, due to the interaction with the W-states (see Fig.~3). However at the Fermi level the $q$-values are larger and fall in the W gap-region; thus the states are localized. In contrast to the majority-spin channel, surface states or resonant states can occur in the minority-spin channel but disappear below the Fermi energy, where the experimental measurements are performed (see Supplementary Note 3 and Supplementary Figure 2).

The flat behavior in the Fermi surface of the localized states in the surface Brillouin zone (upper panel of Fig.~3) can be qualitatively understood from the three-dimensional Fermi surface of the majority-spin bands of bulk Fe. If we consider a finite, but large number N of (001) Fe layers in the vacuum, the $k_{z}$-values are basically quantized into N values and the two-dimensional Fermi surface is a projection of N slices of the 3D Fermi surface. The majority-spin Fermi surface of Fe has nearly planar areas in $\overline\Gamma \overline {\mathrm M} $ directions, such that the sliced Fermi surface sheets result upon projection in nearly straight lines in the gap-areas. This trend can also be seen from Fig.~2 (c), where the two-dimensional Fermi surface is the outcome of the projection of about 200 slices of the 3D Fermi surface.

\subsection*{Friedel Oscillations} In order to understand the STM experiments for Fe layers on W(001), we have to discuss the Friedel oscillations around impurities in metallic systems. Using the stationary phase approximation, for large  distances $R$ from the impurities the energy dependent charge density $\Delta n{(\mathbf{r}+\mathbf{R};E)}$ 
around the Fermi surface is determined from points $\mathbf{k}_{i}$ on the Fermi surface, for which the group velocity $\mathbf{v}_{k}$ points in the directions of $\mathbf{R}$.  If for a given $\mathbf{R}$ only one such $\mathbf{k}$-value, $\mathbf{k}_{1}$, exists on the constant energy surface ($E$=const.), then in three dimensions the energy dependent charge density $\Delta n{(\mathbf{r}+\mathbf{R};E)}$ induced at a large distance $R$ from the impurity is given by ~\cite{Lounis} 

\begin{equation}
\Delta n(\mathbf{r}+\mathbf{R}; E) \propto |\Psi_{\mathbf{k}_{1}}{(\mathbf{r})}|^2|t_{\mathbf{k}_1;-\mathbf{k}_{1}}| \frac{\sin{(2k_{1}^{\perp}R+2\varphi_1+\delta_{1})}}
{R^{2}\,C_1}
\end{equation}
where $C_1={\bigg|\frac{\partial^2E}{\partial k_{x1}^2}\frac{\partial^2E}{\partial k_{y1}^2}\bigg|}$ is the curvature  at point $\mathbf{k}_1$ measuring the flatness of the constant-energy surface, while $k_{1}^{\perp}$ is the component of $\mathbf{k}_1$ perpendicular to the constant-energy surface. $t_{\mathbf{k}_1;-\mathbf{k}_{1}}$, $\delta$ and $\varphi$ are energy dependent quantities defining respectively, 
the scattering strength, the phase-shift that Bloch wave functions experience after scattering and a contour related phase.

Thus, if the curvature $C_1$ at the surface point $\mathbf{k}_1$ is small, the density change in the direction perpendicular to the constant-energy surface is particularly large resulting from the focusing of the group velocities in this direction. In fact, if $C_1$ vanishes, the charge density decays in this direction with a power smaller than 2. If the curvature vanishes in a region larger than a single point, then the changes are even stronger: $\Delta n{(\mathbf{r}+\mathbf{R};E)}$ decays either as $1/R$ if $C_1$ vanishes along a line, or it does not decay at all, if $C_1$ vanishes in a planar area of the constant-energy surface. As discussed in~\cite{Lounis}, often more than one critical point with group velocities $\mathbf{v}_{\mathbf{k}}$ parallel to $\mathbf{R}$ exist, say $\mathbf{k}_1$ and $\mathbf{k}_2$; then the contributions from these two critical points are not just additive, but interfere with each other, so that three oscillation periods determined by $2k_1$, $2k_2$ and $k_1+k_2$ occur. The behavior of Friedel oscillations in two dimensions are very similar as e.g. given by equation (1), with the modification that the amplitude decays as $1/R$ only and $C_1$ is given by the second derivative of the constant energy line.

Coming back to the calculated 2D Fermi surfaces in Fig.~3 we can connect the QWS with the Friedel oscillation. The group velocities of all QWS point into the [110] direction, so we expect very strong charge density oscillations in these directions, as found for the 3 MLs case by von Bergmann et al. (Fig.~1 (b))~\cite{Bergmann} and which is in agreement with our results in Fig.~1 (d). Clearly only the 
majority-spin Fe states give rise to this, since the minority electrons do not have quantum well states in the gap region (see Fig.~2 (b)). Fig.~4 shows the calculated Friedel oscillations along the [110] direction, for 3 MLs, 5 MLs, 7 MLs and 11 MLs of Fe. Firstly we see that the oscillations for 3 MLs are strong and very long ranged, arising from the fact that (i) the oscillations arise from a quantum well state and that (ii) the curvature of this state is very small. Moreover, due to the single QWS, the oscillations reflect the periodicity determined by the different $k_i^{\perp}$ found on parallel flat parts of the Fermi surface. The behavior for 7 and 11 MLs shows a multiperiodicity arising from the 3 respectively 5 quantum well states. This leads to very large oscillation amplitudes, but due to partly destructive interferences also to an additional modulation structure with periods determined by the distance $\Delta k$ between the 3, 5, 7, and 11 different QWS. For thicker films these nodes shift to larger distances, as one would expect from the decrease of $\Delta k$ and increase of the film thickness. In general we expect that these oscillations, observed for 3 MLs Fe up to 5 nm, should be detectable for thicker films to up an order of magnitude larger distances. For 2 MLs, investigated experimentally~\cite{Bergmann}, the QWS does not cross the Fermi level and does not contribute to the Friedel oscillations measured at the Fermi energy. This explains the extinction of the charge ripples at large distances. 

Concerning the minority-spin channel, we found that surface states and resonant states can occur which leads necessarily to charge oscillations. However, for 3 MLs the minority-spin surface states 
disappear below the Fermi energy 
(see Supplementary Note 3 and Supplementary Figure 2) and thus do not contribute to the charge density oscillations. Above the Fermi energy, 
oscillations are obtained but their decay is surprisingly much stronger than those obtained from the QWS of 
the majority-spin channel.

\section*{Discussions}
We find large focused long-range and large amplitude majority-spin density oscillations at the Fe surface created by spin-polarized quantum well states 
of Fe films on W(100).  Our finding supports the possibility to design and control of quantum well states in supported thin layers that lead to a super-amplification of interaction effects. 
This may open new avenues of research by cross fertilizing communities working on quantum well states and on surface magnetism through the investigation of the  
charge or spin density oscillations along the directions parallel to the films. 
This can impact on the lateral coupling of magnetic nanostructures at the vicinity of the surface. For magnetic films, in particular, there is a strong amplification of a selected spin component 
(spin filtering) which can be detected at large distances from embedded non-magnetic defects. This may have an impact in nanotechnology, since spin-information can be transmitted laterally to 
large distances. The choice of the substrate and the thickness of the  film are  of crucial  importance in  terms  of their electronic structure to obtain a giant laterally  focused signal 
far away from the  imperfections located in the surface of thin films. Many such systems should exist, making charge density oscillations in thin films of stronger importance than in bulk materials. 
As substrates the (001) surfaces of W, Cr and Mo are good choices, but also the semiconductor surfaces should be interesting candidates. For the film materials, metals with relatively flat Fermi 
surface parts, like ferromagnetic Fe, Co and Ni, but also Pd and Pt and even the noble metals ~\cite{Zahn} would be potential materials to create long ranged and strongly anisotropic charge 
density oscillations within the film.

\section*{Methods}
\subsection*{First-principles simulations} 
The calculations are based on DFT. First, the atomic positions of Fe 
atoms are relaxed using the full-potential linearized augmented plane wave (FLAPW) method within DFT as implemented in the FLEUR package~\cite{fleur1,fleur2}.
The relaxations favor inwards displacements of Fe atoms along the z axis, i.e. towards the substrate.
Once the positions are known (see Supplementary Note 1 and Supplementary Table 1), the same type of simulations are repeated with the Korringa-Kohn-Rostoker 
Green function method~\cite{Papanikolaou} in its full-potential version considering a semi-infinite W(001). The latter 
allows to avoid size effects using the decimation technique~\cite{Moliner,Szunyogh}. Once the surface is simulated, the impurity is considered and 
the embedding technique based on Green functions is used to evaluate the change in the charge density at different locations.

\section*{Acknowledgments}
We would like to thank Kirsten von Bergmann for pointing out her measurements, for discussions and for providing their experimental data. This work was funded by the HGF-YIG programme Funsilab--Functional Nanoscale Structure Probe and Simulation Laboratory (VH-NG-717). B. Z. is supported by the HGF-YIG programme VH-NG-513. We acknowledge computing time on the supercomputers JUQUEEN, JUROPA at J\"ulich Supercomputing Centre and JARA-HPC of RWTH Aachen University.

\section*{Author contributions}
M.B. and S.L. performed and analyzed the ab-initio calculations. M.B, P.H.D. and S.L. wrote the manuscript. All authors discussed the results and commented on the manuscript.

\section*{Competing financial interests}
The authors declare no competing financial interests.
\newpage
\begin{figure}
\begin{center}
\includegraphics*[angle=0,width=1\linewidth]{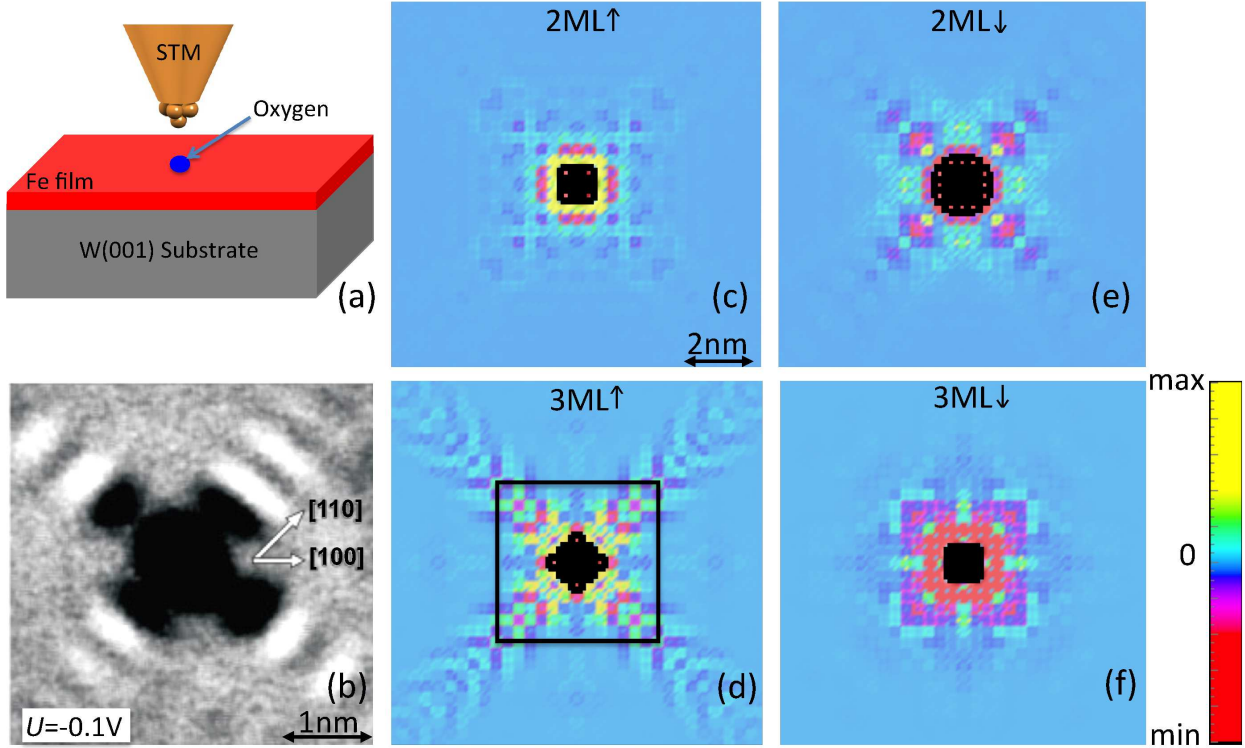}
\end{center}
\caption{{\bf Charge density oscillations induced around an impurity on thin Fe films on W(001)}. (a) STM tip scanning the Fe film surface where the Oxygen impurity is implanted. The Fe film is grown on the (001) surface of W. (b) Measured charge density of states (d$I$/d$U$ map) using a W-tip at the vicinity of the Fermi energy ($U$ = -0.1 eV) 
induced by an oxygen impurity on 3 MLs Fe/W(001)~\cite{Bergmann} where the bright (dark) color stands for charge density accumulation (depletion).
Comparison to the spin-resolved theoretical charge density oscillations for states at the Fermi energy  emanating from an oxygen impurity 
for two film thicknesses: 2 MLs and 3 MLs for the majority-spin channel (c) and 
(d) followed by the corresponding ones for the minority-spin channel (e) and (f). The square in (d) encompasses the region probed experimentally and the black spots are obtained after an uppercut of the theoretical data.}
\label{combination1}
\end{figure}

\begin{figure}
\begin{center}
\includegraphics*[width=\linewidth]{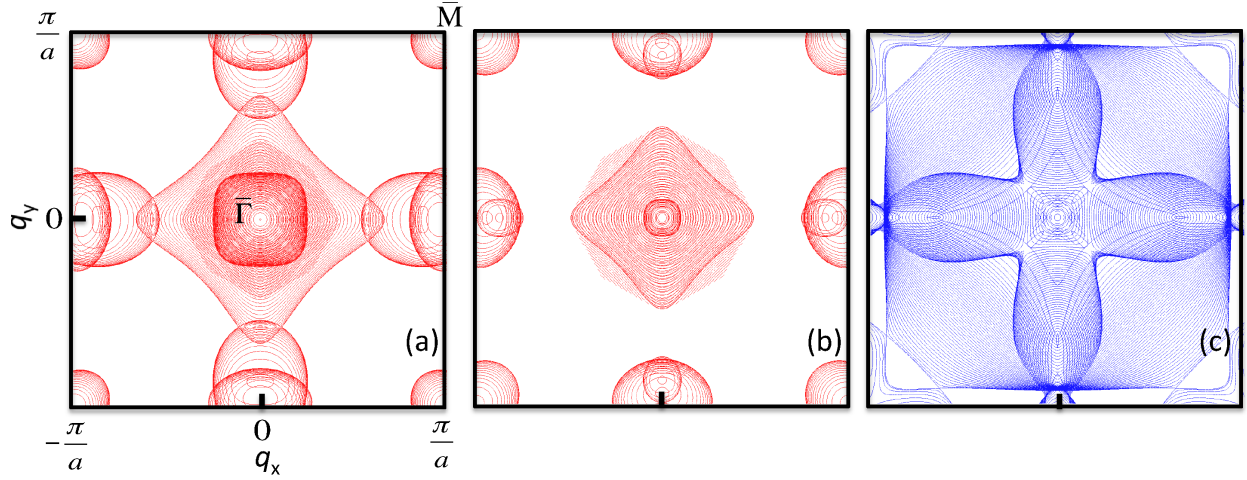}
\end{center}
\caption{{\bf Projected Bloch spectral function of bulk W and Fe at the Fermi energy plotted in the reciprocal space.} The bulk Fermi surface is projected onto the (001) surface Brillouin zone for (a) W, (b) Fe minority-spin channel and (c) Fe majority-spin channel.}
\label{combination4}
\end{figure}

\begin{figure}
\begin{center}
\includegraphics*[width=\linewidth]{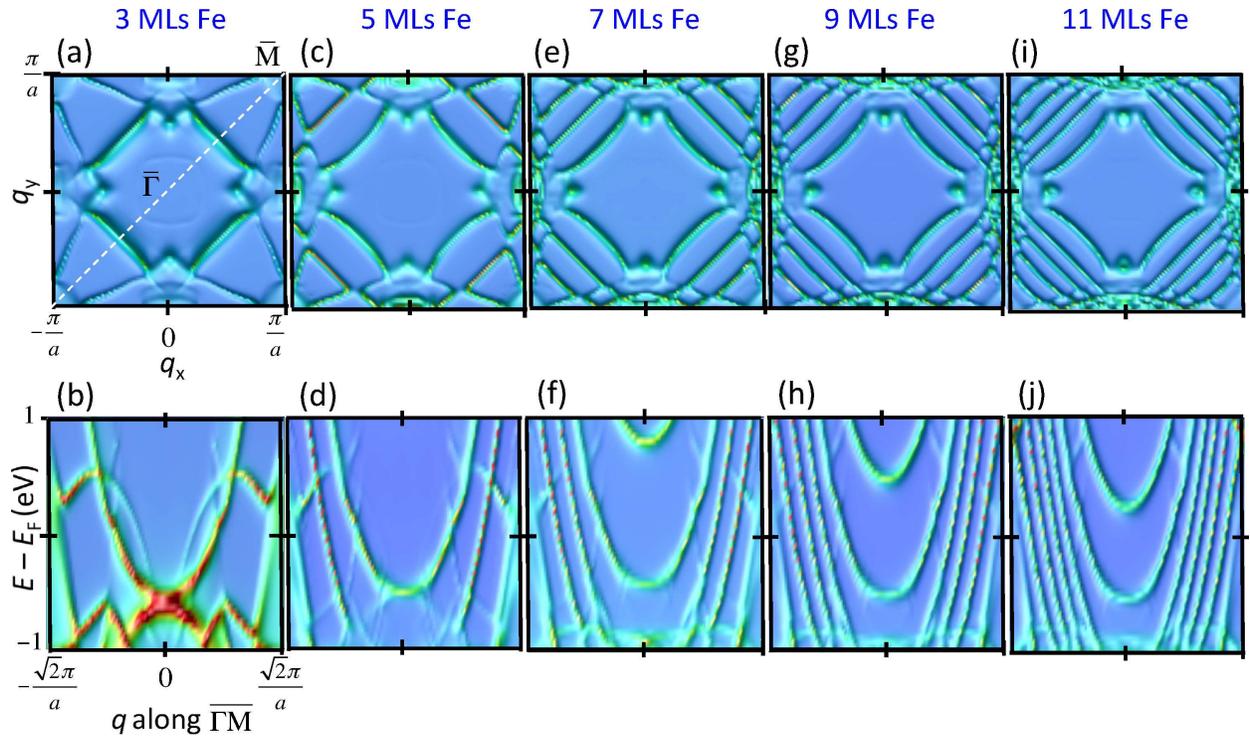}
\end{center}
\caption{{\bf Quantum well states in the majority-spin channels of Fe films/W(001) surface.} 
The upper figures exhibit the two-dimensional Bloch spectral function in the first Brillouin zone of the majority-spin channel of Fe surface atoms at the Fermi energy. Results for five different film thicknesses are depicted, i.e.  3, 5, 7, 9 and 11 Fe MLs on W(001). In the lower pictures, the energy dependence of the spectral function along the $\overline\Gamma \overline{\mathrm M}$ direction (dashed-line in upper left) is plotted for the respective film thicknesses.}
\label{combination3}
\end{figure}

\begin{figure}
\begin{center}
\includegraphics*[width=\linewidth]{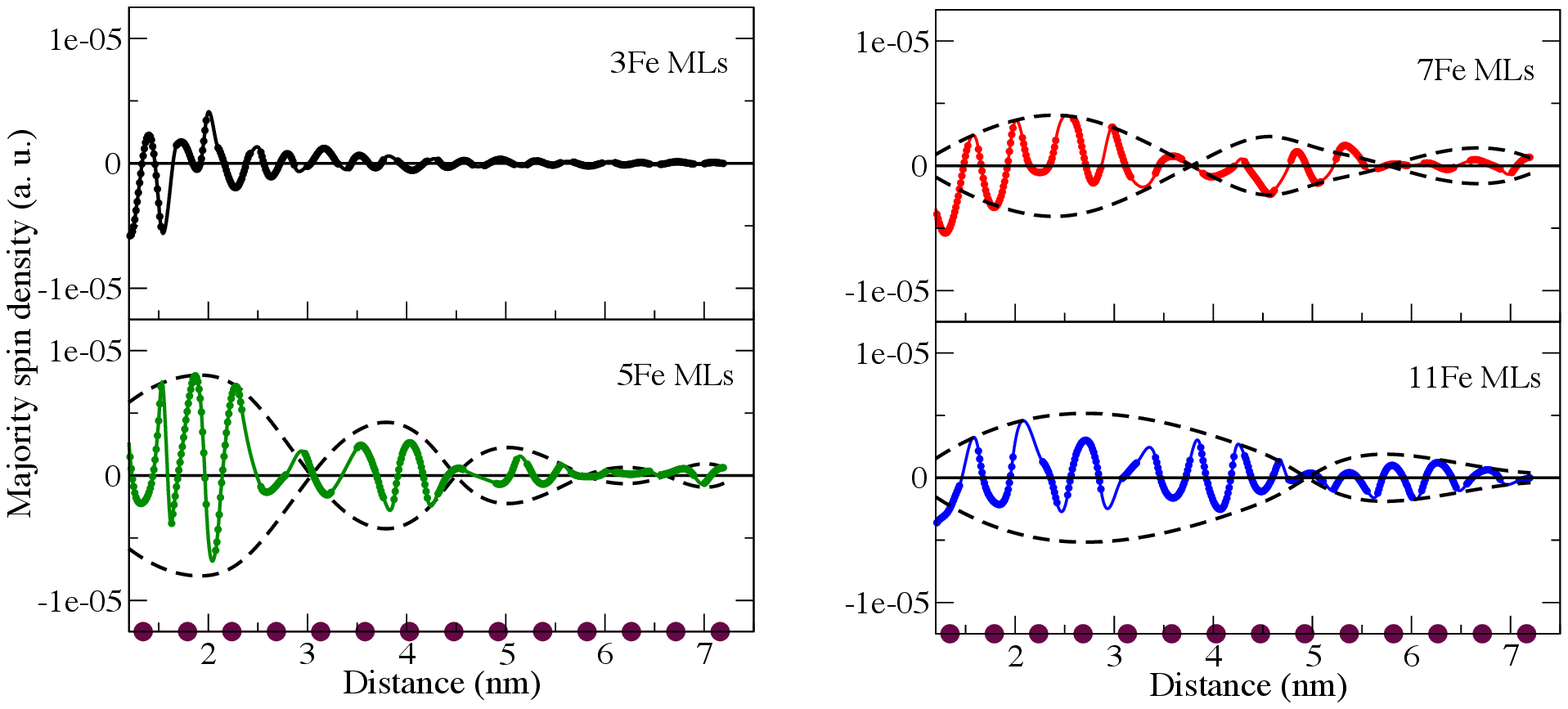}
\end{center}
\caption{{\bf Theoretical induced majority-spin density waves in Fe films on W(001) surface.} (a) Comparison of the majority-spin densities induced by oxygen impurity implanted in the (001) surface of Fe films with thickness of 3, 5, 7 and 11MLs. (b) Dashed lines showing different periods of higher amplitudes of the induced charge density along the surface. The dashed lines are added only to guide the eye.}
\label{combination3ab}
\end{figure}


\begin{thebibliography}{99}
\bibitem{Crommie1}{Crommie, M. F., Lutz, C. P., \& Eigler, D. M. Imaging standing waves in a two-dimensional electron gas. Nature \textbf{363}, 524-527 (1993)} 
\bibitem{Avouris1}{Hasegawa, Y., \& Avouris, Ph. Direct Observation of standing wave formation at surface steps using scanning tunneling spectroscopy. Phys. Rev. Lett. \textbf{71}, 1071-1074 (1993)}
\bibitem{Crommie2}{Crommie, M. F., Lutz, C. P., \& Eigler, D. M. Confinement of electrons to quantum corrals on a metal surface. Science \textbf{262}, 218-220 (1993)} 
\bibitem{Silly}{Silly, F., Pivetta, M., Ternes, M., Patthey, F., Pelz, J. P., Schneider, W.-D. Creation of an atomic superlattice by immersing metallic adaptors in a two-dimensional electron sea. Phys. Rev. Lett. \textbf{92}, 16101 (2004)}
\bibitem{Lounis2}{Lounis, S., Bringer, A., Bl\"ugel, S. Magnetic adatom induced skyrmion-like spin texture in surface electron waves. Phys. Rev. Lett. {\bf 108}, 207202 (2012)}
\bibitem{Meier}{Meier, F., Lounis, S., Wiebe, J., Zhou, L., Heers, S., Mavropoulos, Ph., Dederichs, P. H., Bl\"ugel, S. 
\& Wiesendanger, R. Spin polarization of platinum (111) induced by the proximity to cobalt nanostripes. Phys. Rev. B 
{\bf 83}, 075407 (2011)}
\bibitem{Weismann}{Weismann, A., Wenderoth, M., Lounis, S., Zahn, P., Quaas, N., Ulbrich, R. G., Dederichs, P. H. \& Bl\"ugel, S. Seeing the Fermi surface in real space by nanoscal electron focusing. Science \textbf{323}, 1190-1193 (2009)}
\bibitem{Lounis}{Lounis, S., Zahn P., Weismann A., Wenderoth, M., Ulbrich, R. G., Mertig, I., Dederichs, P. H. \& Bl\"ugel, S. Theory of real space imaging of Fermi surface parts. Phys. Rev. B      \textbf{83}, 35427 (2011)}
\bibitem{Avotina} {Avotina, Ye. S., Kolesnichenko, Y. A., Otte, A. F. \& van Ruitenbeek. Signature of Fermi surface anisotropy in point contact conductance in the presence 
of defects. Phys. Rev. B \textbf{74}, 085411 (2006)}
\bibitem{Guo}{Guo, Y., Zhang, Y.-F., Bao, X.-Y., Han, T.-Z., Tang, Z., Zhang, L.-X., Zhu, W.-G., Wang, E. G., 
Niu, Q., Qiu, Z. Q., Jia, J.-F., Zhao, Z.-X., \& Xue, Q.-K. Superconductivity modulated by quantum size effects. 
Science \textbf{306}, 1915-1917 (2004)}
\bibitem{Ma}{Ma, X., Jiang, P., Qi, Y., Jia, J., Yang, Y., Duan, W., Li, W.-X., Bao, X., Zhang, S. B., \& Xue, Q.-K. 
Experimental observation of quantum oscillation of surface chemical reactivities. PNAS \textbf{104}, 9204-9208 (2007)}
\bibitem{Yang}{Yang, H., Peng, X., Wei, X., Liu, W., Zhu, W., Xiao, D., Stocks, G. M., \& Zhong, J. Quantum 
oscillation of Rashba spin splitting in topological insulator Bi$_2$Se$_3$ induced by the quantum size effects 
of Pb adlayers. Phys. Rev. B \textbf{86}, 155317 (2012)}
\bibitem{Grunberg}{Gr\"unbeg, P., Schreiber, R., Pang, Y., Brodsky, M. B. \& Sowers, H. Layered Magnetic Structures: Evidence for antiferromagnetic coupling of Fe layers across Cr interlayers. Phys. Rev. Lett. \textbf{57}, 2442-2445 (1986)}
\bibitem{Baibich}{Baibich, M. N. et al. Giant magnetoresistance of (001)Fe/(001)Cr magnetic superlattices. Phys. Rev.Lett. \textbf{61}, 2472-2475 (1988)}
\bibitem{Parkin}{Parkin, S. S. P., More, N. \& Roche, K. P. Oscillations in exchange coupling and magnetoresistance in metallic superlattice structures: Co/Ru, Co/Cr, and Fe/Cr. Phys. Rev. Lett. \textbf{64}, 2304-2307 (1990)}
\bibitem{Ortega}{Ortega, J. E. \& Himpsel, F. J. Quantum well states as mediators of magnetic coupling in superlattices. Phys. Rev. Lett. \textbf{69}, 844-847 (1992)}
\bibitem{Carbone}{Carbone, C., Vescovo, E., Rader, O., Gudat, W. \& Eberhardt, W. Exchange split quantum well states of a noble metal film on a magnetic substrate. Phys. Rev. Lett. \textbf{71}, 2805-2808 (1993)}
 \bibitem{Zhou}{Zhou, L., Wiebe, J., Lounis, S., Vedmedenko, E., Meier, F., Bl\"ugel, S., Dederichs, P. H. \& Wiesendanger, R. Strength and directionality of surface Ruderman-Kittel-Kasuya-Yosida interaction mapped on the atomic scale. Nature Physics, {\bf 6}, 187-191 (2010)}
\bibitem{Khajetoorians1}{Khajetoorians, A. A., Wiebe, J., Chilian, B., Lounis, S., Bl\"ugel, S. \& Wiesendanger, R. Atom-by-atom engineering and magnetometry of tailored nanomagnets. Nature Physics, {\bf 8}, 497-503 (2012)} 
\bibitem{Khajetoorians2}{Khajetoorians, A. A., Wiebe, J., Chilian, B. \& Wiesendanger, R. Realizing all-spin-based logic operations atom by atom. Science {\bf 332}, 1062-1064 (2011)}
\bibitem{Ngo}{Ngo, A. T., Rodriguez-Laguna J., Ulloa, E. S., \& Kim, H. E. Quantum manipulation via atomic-scale magnetoelectric effects. Nano Lett. {\bf 12}, 13-16 (2012)}
\bibitem{Pruser}{Pr\"user, H., Dargel, P. E., Bouhassoune, M., Ulbrich, R. G., Pruschke, T., Lounis, S. \& Wenderoth, M. Interplay between Kondo effect and Ruderman-Kittel-Kasuya-Yosida interaction. Nature Communications, accepted (2014)}
\bibitem{Stepanyuk1} {Stepanyuk, V. S., Negulyaev, N. N., Niebergall, L. \& Bruno, P. Effect of quantum confinement of surface electrons on adatom-adatom interactions. New J. Phys. {\bf 9}, 388 (2007)}
\bibitem{Stepanyuk2}{Brovko, O. O., Ignatiev, P. A., Stepanyuk, V. S. \& Bruno, P. Tailoring exchange interactions in engineered nanostructures: An ab initio study. Phys. Rev. Lett. {\bf 101}, 036809 (2008)}
 \bibitem{Bergmann}{von Bergmann, K. Iron nanostructures studied by spin-polarised scanning tunneling microscopy. PhD Thesis, University of Hamburg (2004)}.
 \bibitem{Spisak}{Spisak, D. \& Hafner, J. Diffusion of Fe atoms on W surfaces and Fe/films and along surface steps. Phys. Rev. B {\bf 70}, 195426 (2004)}
\bibitem{Kubetzka}{Kubetzka, A., Ferriani, P., Bode, M., Heinze, S., Bihlmayer, G., von Bergmann, K., Pietzsch, O., Bl\"ugel, S., \& Wiesendanger, R. 
Revealing antiferromagnetic order of the Fe monolayer, on W(001): Spin-polarized scanning tunneling microscopy and first-principles calculations. Phys. Rev. Lett. {\bf 94}, 087204 (2005)} 
\bibitem{vonBergmann_PRB}{von Bergmann, K. Bode, M. Wiesendanger, R. Magnetism of iron on tungsten (001) studied by spin-resolved scanning tunneling microscopy and spectroscopy. Phys. Rev. B {\bf 70}, 174455 (2004)}
\bibitem{Zahn}{http://www.hzdr.de/projects/fermisur/}
\bibitem{fleur1}{http://www.flapw.de}
\bibitem{fleur2}{Kurz, Ph. F\"orster, F. Nordstr\"om, L., Bihlmayer, G. and Bl\"ugel, S. Ab-initio treatment of noncollinear magnets with the 
full-potential linearized augmented plane wave method. Phys. Rev. B {\bf 69}, 024415 (2004)}
\bibitem{Papanikolaou}{Papanikolaou, N., Zeller, R. \& Dederichs, P. H. Conceptual improvements of the KKR method. J. Phys. Condens. Matter \textbf{14}, 2799-2823 (2002)} 
\bibitem{Moliner}{Garcia-Moliner, F. \& Velasco, V. R. Theory of incomplete crystals, surfaces, defects, interfaces, and layered structures. Progr. Surf. Sci. \textbf{21}, 93-162 (1986)}
 \bibitem{Szunyogh}{Szunyogh, L., Ujfalussy, B., Weinberger, P. \& Kollar, J. Self-consistent localized KKR scheme for surfaces and interfaces. Phys. Rev. B \textbf{49}, 2721-2729 (1994)}

 
\end{thebibliography}
\end{document}